\documentclass{article}
\usepackage[utf8]{inputenc}
\usepackage{amsmath,amsthm,amssymb,rotating,hyperref,booktabs}
\usepackage[all,cmtip]{xy}
\usepackage[osf]{mathpazo}

\theoremstyle{definition}
\newtheorem{example}{Example}
\newtheorem{defn}{Definition}
\DeclareMathOperator\CE{CE}
\DeclareMathOperator\Weil{Weil}
\title{AKSZ-type Topological Quantum Field Theories and Rational Homotopy Theory}
\author{Hyungrok Kim\footnote{Department of Physics, Stony Brook University}}
\begin{document}
\maketitle

\begin{abstract}
We reformulate and motivate AKSZ-type topological field theories in pedestrian terms, explaining how they arise as the most general Schwartz-type topological actions subject to a simple constraint, and how they generalize Chern--Simons theory and other well known topological field theories, in that they are gauge theories of flat connections of higher gauge groups (infinity-Lie algebras).

Their Euler--Lagrange equations define quasifree graded-commutative differential algebras, or equivalently \(L_\infty\)-algebras, the equivalent of the Lie algebra of the gauge group; we explain how integrating out auxiliary fields in physics corresponds to taking the Sullivan minimal model of this algebra, and how the correspondence between fields and gauge transformations realizes Koszul duality.

Using this dictionary, we can import topological invariants and notions (e.g.~the rational LS-category) to apply to this class of theories.

\end{abstract}

\tableofcontents
\section{Introduction}
It is by now well accepted that particular classes of physical theories correspond to particular classes of geometries: e.g. a classical particle can live on a Riemannian manifold; more complicated physical models (e.g.~including external electromagnetic fields) correspond to additional geometric structure on this manifold; a quantum particle sees the Laplacian spectrum instead; a classical or quantum string sees a yet different structure; certain superconformal field theories can be thought of as sigma models on the affine schemes defined by their chiral rings; vertex- or edge-weighted graphs correspond to Ising models with inhomogeneous magnetic field or temperature; 1-factorable graphs to dimer models; and so on.

In this work we consider a class of physical theories, the AKSZ-type TQFT, that represent spaces up to a very weak notion of equivalence (real homotopy equivalence), and equipped with a generalization of an invariant binary form. The study of these topological structures is called \emph{rational homotopy theory}, initiated by Sullivan. AKSZ-type theories are a class of topological field theories with differential-form fields (including scalars and Yang–Mills fields), originally proposed in \cite{aksz}, motivated by considerations of BRST quantization and mirror symmetry; later deep relations with higher category theory and homotopy theory were found (see e.g.~references in \cite{frs}).

In this work, we explain how various constructions in rational homotopy theory, including Koszul duality and Sullivan resolution, correspond to natural operations on these physical theories, and conversely how homotopic invariants define interesting invariants of these physical theories.

The structure of gauge symmetries in this class of theories is richer than the typical case: while (in the minimal case) they form a Lie supergroup as usual, in addition they have higher-order operations, corresponding to higher-order brackets in an \(L_\infty\)-algebra.

There are two central dichotomies that we wish to highlight in the class of topological field theories in question: simply connected vs.~multiply connected and minimal vs.~nonminimal. Simply connectedness means, in topology, that the degree-1 and degree-0 homotopy groups be trivial; the corresponding feature in physics is that fields of differential-form degrees 0 and 1 --- that is, the scalars and the Yang--Mills fields --- be absent. In topology, the failure of simply-connectedness is manifest in e.g.~the failure of ratinoal homotopy theory to work well for such spaces;\footnote{In this subject, one typically requires several simplifying assumptions on these low-degree homotopy groups, viz.~that the space in question be \emph{simple} or \emph{nilpotent} \cite{fht}.} in physics, this manifests as an essential non-Abelianness of the theories in question: higher-degree gauge fields of degree higher than that of the differential are, in a certain sense, Abelian, due to mere degree reasons.

The other dichotomy that we wish to highlight is that of minimality. The terminology is borrowed from rational homotopy theory, where one speaks of minimal or non-minimal Sullival algebras; essentially, non-minimal theories are those containing a “mass” (i.e.~quadratic) term in their potentials, while in non-minimal theories these terms are absent. The presence of such terms renders some of the fields in the action to be auxiliary; their function is similar to that of the auxiliary fields \(B\) occurring in the Mathai--Quillen formalism for cohomological (Witten-type) topological field theories in an action of the form
\[\int B F^+ + \frac12 B^2\]
(see \cite[Chapter~6]{lm} for a pedagogical discussion). Being auxiliary, these fields can be integrated out and, in the simply connected case, leaves another AKSZ-type theory without the mass terms. Their elimination corresponds to the minimization of a non-minimal Sullivan algebra into a minimal one in topology, which essentially amounts to eliminating, in a cellular complex, filled balls, which do not affect the homotopy type of a space. Complications arise if both multiply-connectedness and non-minimality are present simultaneously: in that case, integrating out the auxiliary fields in general leaves the action in a non-AKSZ form.

\subsection{Conventions}
All coefficient fields will be that of the real numbers, unless otherwise specified: we regard algebra generators as fields, and these are real-valued in field theory in the physics sense. From a topological standpoint, rational homotopy theory, as its name implies, is usually done over the field of rational numbers; its discreteness mirrors that of the homotopy classes of topological spaces: that is, the space of homotopy classes of topological spaces is discrete, and does not have real moduli; and, if we forget torsion (and assume technical conditions about connectedness and simply-connectedness) these correspond to certain algebras over rationals. Working over the reals instead of the rationals, therefore, has the effect of continuously “interpolating” between different homotopy types, a concept bizarre to a topologist but more familiar to physicists.

We also neglect all super-phenomena.\footnote{Except in the trivial sense that we can quotient the \(\mathbb N\)-grading into \(\mathbb Z/(2)\), i.e.~that odd-degree generators anticommute.}. In physics terms, this means all fields are bosonic; in mathematical terms, this means we work with Lie algebras rather than Lie superalgebras, manifolds rather than supermanifolds, etc., and that things are \(\mathbb N\)-graded rather than \(\mathbb N\oplus\mathbb Z/(2)\)-graded. Almost everything here can be super-ized in an obvious way, however, by throwing in additional minus signs appropriately.

\subsection{Acknowledgements}
I would like to thank Martin Roček for his highly helpful discussions, unflagging support, and for initial exploratory ideas on which this paper is based.

I would also like to thank Ingmar Saberi for his invaluable comments and suggestions, as well as Dennis Sullivan for helpful discussions.

\section{Review of AKSZ-type theories}
As much of the literature \cite{aksz,frs,ikeda,roytenberg} \cite[Letter 8]{severa00} of this class of Schwartz-type TQFTs is couched or motivated in terms of either categorical homotopy theory (“\(\infty\)-Chern–Weil theory”) or Batalin–Vilkovisky quantization, we find it worthwhile to write down, in plain language, the construction of this class of thoeries as merely Schwartz–type TQFTs.

The procedure boils down to these steps.
\begin{enumerate}
\item Pick your favourite worldvolume dimension \(n\).
\item Put in pairs of differential-form-valued fields, such that their degrees sum to \(n-1\). For example, if a 2-form field exists, then a corresponding \(n-3\)-form field exists also. (A 0-form is exceptional in that it will in general live on a target manifold, making the theory a topological \(\sigma\)-model.) Let the fields be \(\phi^i\); using this pairing we can raise and lower this index:
\[\deg\phi^i + \deg\phi_i = n-1.\]
\item Write down the topological “kinetic term” \(\phi_i\wedge\mathrm d\phi^i\).
\item Write down the most general possible “potential term” in terms of wedge products of differerential forms. There will be a finite number of possible terms unless a 0-form exists (in which case one will have an arbitrary smooth function of the scalar form). That is,
\[
S=k\left(\phi_i\wedge\mathrm d\phi^i + p(\phi_1,\phi_2,\dotsc)\right)
\]
where \(p\) is a polynomial of wedge products of differential forms whose terms are homogeneous in degree \(n\), at least if scalars are absent. Here \(k\in\mathbb R\) is an overall constant that is irrelevant in the classical theory.
\item The most general potential \(p\) involves a number of structure constants. One must solve for \emph{generalized Jacobi identities} satisfied by these, which are required for invariance under small gauge transformations.
\item The theory naturally comes with a set of gauge symmetries, which generalize that of Abelian \(p\)-form fields and Yang–Mills fields. Every field (other than possibly scalars) will have an associated gauge symmetry.\footnote{For scalars, see the discussion in Example~\ref{2d-gauge-discussion}.}
\item The structure thus produced can be regarded as a flat connection for a flat \emph{gerbe} or \emph{principal bundle} (for an \(L_\infty\)-algebra); one can introduce nontrivial boundary conditions instead, producing topologically nontrivial gerbes or bundles.\footnote{Note that this is a very different notion than nontrivial flat connections on a (possibly flat) topologically trivial bundles.} That is, one only requires that the fields be defined patchwise with respect to an open cover, and allow gauge transformatinos as transition maps, with appropriate cocycle conditions (cf. the discussion in \cite{frs}).\footnote{This step requires finite, exponentiated gauge transformations; therefore one must choose the gauge \(\infty\)-group corresponding to the gauge algebra, e.g.~between \(\operatorname{SO}(3)\) and \(\operatorname{SU}(2)\) for the algebra \(\mathfrak{so}(3)\cong\mathfrak{su}(2)\).}
\item In the quantum theory, the overall coefficient \(k\) of this action may be quantized due to the requirement that the action change by \(2\pi\mathbb Z\) under large gauge transformations.
\end{enumerate}

It will be illustrative to look at a few examples.

\begin{example}[Particle on symplectic manifold]
In one dimension, we can only have 0-forms (scalars). Let the fields be \(x^i\), to be raised and lowered by constants \(\omega_{ij}\). Thus, the general action is
\begin{equation}
S = k\int_{\mathbb R} \omega_{ij}\phi^i\mathrm d\phi^j.
\end{equation}
No potential terms are possible. Using integration by parts, we can arrange that \(\omega_{ij}=-\omega_{ji}\). Thus \(\omega\) is (locally) a nondegenerate antisymmetric matrix.

The scalars live on a manifold \(M\). Locally, the coefficients \(\omega_{ij}\) are constants; this means that \((M,\omega)\) is globally a symplectic matrix, by Darboux’s theorem. That is, abstractly regarding \(\phi\) as defining a curve
\[\phi\colon \mathbb R\to M,\]
the action is
\[S = k\int\phi^*\alpha,\]
where \(\alpha\) is a locally defined antiderivative of the symplectic form:
\[\mathrm d\alpha = \omega.\]
That is, for a worldvolume that is a closed curve \(\gamma=\partial\Sigma\) that is the boundary of a surface \(\Sigma\subseteq M\), the action is
\begin{equation}
S=k \iint_\Sigma \omega.
\end{equation}
Now, quantization requires that \(\exp(2\pi\mathrm iS)\) be well defined regardless of the choice of the surface \(\Sigma\) that bounds the closed worldline \(\gamma\). If we normalize (without loss of generality) the de~Rham cohomology class of \(\omega\) to lie in integral cohomology (modulo torsion),
\begin{equation}
[\omega] \in \operatorname H^2(M;\mathbb Z)/\operatorname{Tors}(\operatorname H^2(M;\mathbb Z)),
\end{equation}
that is, require that \(\oint_\Sigma\omega\in\mathbb Z\) for every closed surface \(\Sigma\subseteq M\), then \(k\in\mathbb 2\pi\mathbb Z\) in order for the quantum theory to be well defined.

The equations of motion of this theory are simply
\[0 = \mathrm d\phi^i \in \Omega^1(\mathbb R;\phi^*\mathrm TM).\]
That is, classically, this describes a particle that remains constant on a symplectic manifold. Explicitly, consider the case where \(M=\mathrm T^*N\) is the cotangent bundle over a manifold \(N\). Then the action is
\[S = k \int p_i \dot q^i\,\mathrm dt,\]
(in this case an antiderivative of the symplectic form exists globally)
and it can be recognized as the infinite-mass limit of the standard action for a first-quantized particle
\[S = k \int \left(p_i \dot q^i - \frac1{2m}g^{ij}p_ip_j\right)\,\mathrm dt\]
where \(g\) is a Riemannian metic on \(M\).
\end{example}

\begin{example}[String on Poisson manifold]
When the worldvolume \(\Sigma\) is two-dimensional, we have a scalar \(x^i\) and a one-form \(e_i\). The action is
\begin{equation}\label{2d-action}
S = 2\pi k\int_\Sigma \left(e_i\wedge\mathrm dx^i - \frac12\pi^{ij}(x)e_i\wedge e_j\right)
\end{equation}
where the function \(\pi^{ij}\) defining the potential is necessarily antisymmetric. Once again, the scalar \(x\) lives on a manifold \(M\); then \(e_i\in \mathrm T^*_xM\) is a covector at the point \(x\in M\), and \(\pi^{ij}\) defines an antisymmetric (2,0)-tensor on \(M\). That is, the data defines a sigma model
\begin{align*}
(x,e) \colon \Sigma &\to \mathrm T^*M\\
z & \mapsto (x(z),e(z))
\end{align*}
on the cotangent bundle of a manifold \(M\). By rescaling \(\pi\mapsto k\pi\) and \(e\mapsto k^{-1}e\) (dimensional analysis), the constant \(k\) can be absorbed into \(\pi\).

The equations of motion are
\begin{align}\label{2d-eom}
\mathrm dx^i &= \pi^{ij}(x)e_j &
\mathrm de_i &= \frac12\partial_i\pi^{jk}(x)e_j\wedge e_k.
\end{align}
For there to be any nontrivial solutions to the equations of motion, the equations of motion must be compatible with the nilpotence of the exterior derivative (i.e.~the Bianchi identity); that is,
\begin{equation}
0 = \mathrm d^2x^i = 
-\frac32e_j\wedge e_k\left(
\pi^{l[i}\partial_l\pi^{jk]}
\right)
\end{equation}
where \([\dotso]\) denotes normalized total antisymmetrization. The bracketed expression can be recognized as the Schouten–Nijnhuis bracket \([\pi,\pi]\); that is, \(\pi\) defines a Poisson structure on \(M\).
\end{example}

\begin{example}[Membrane on a Lie algebra, a.k.a.~Chern–Simons]
On a three-dimensional worldvolume, we can have a scalar/2-form pair or a 1-form/1-form pair. For simplicity let us treat only the 1-form/1-form case. (The general case is known in the literature as the \emph{Courant \(\sigma\)-model} \cite{roytenberg,roytenbergthesis,ikeda}.)

Let the 1-form fields be \(A^i\). Then the action is
\[
S = k\int \frac12 B_{ij}A^i\mathrm dA^j + \frac1{3!}f_{ijk}A^i\wedge A^j\wedge A^k
\]
where \(B_{ij}\) is a symmetric matrix and \(f_{ijk}\) are totally antisymmetric structure constants. The equations of motion are
\begin{equation}
\mathrm dA^i + \frac12f^i{}_{jk}A^j\wedge A^k = 0.
\end{equation}
The consistency condition \(\mathrm d^2A^i = 0\) reduces to the Jacobi identity for \(f\), which therefore defines a Lie algebra. This case is therefore the Chern–Simons theory on a Lie algebra equipped with an invariant nondegenerate bilinear form (not necessarily semisimple). It is well known to exhibit level quantization if this Lie algebra is non-Abelian.
\end{example}

\begin{example}[Four dimensions]
On a four-dimensional worldvolume, again suppressing a scalar field, we have a 1-form \(A^i\) and a 2-form \(B_i\), with the general action
\begin{equation}\label{4d-action}
S = \int B_i \wedge \mathrm dA^i + 
\frac12 \eta^{ij} B_i \wedge B_j +
\frac12 f^i{}_{jk} B_i \wedge A^j \wedge A^k + \frac1{4!}h_{ijkl}A^i\wedge A^j\wedge A^k\wedge A^l.
\end{equation}
The coefficients \(\eta^{ij}\), \(f^i_{jk}\) and \(h_{ijkl}\) satisfy obvious (anti-)symmetries
\begin{align}
\eta^{ij} &= \eta^{ji} &
f^i{}_{jk} &= -f^i{}_{kj} &
h_{ijkl} &= h_{[ijkl]}. 
\end{align}

The equations of motion are
\begin{align}
\mathrm dA^i + \eta^{ij}B_j + \frac12 f^i{}_{jk}A^j \wedge A^k &= 0\\
\mathrm dB_i + f^k{}_{ij}B_k\wedge A^j + \frac16h_{ijkl}A^j\wedge A^k\wedge A^l &=0.
\end{align}
In order for the Bianchi identities \(\mathrm d^2A^i =0 \), \(\mathrm d^2B_i =0\) to hold, the coupling constants \((\eta,f,h)\) must satisfy certain relations.
\begin{itemize}
\item Taking the expression \(\mathrm d^2A\) and leaving only the terms proportional to \(A^3\), we obtain

\begin{equation}\label{4d-jacobi}
\frac13\eta^{ij}h_{jklm}
+
f^i{}_{j[k}
f^j{}_{lm]}
=0.
\end{equation}
That is, the Jacobi identity holds up to a term proportional to \(h\), which can be thought of as the coefficient for a trilinear operation. (Later, we will see that this can be interpreted as a “homotopic Jacobi identity”, where \(\eta\) is interpreted as a differential.) Taking the coefficient of \(A^2B\) in \(\mathrm d^2B\) gives the same result.
\item Taking the expression \(\mathrm d^2B\) and leaving only the terms proportional to \(B^2\), we obtain
\begin{equation}\label{4d-restriction}
\eta^{i(j}f^{k)}{}_{il}= 0 .
\end{equation}
Taking the coefficient of \(AB\) in \(\mathrm d^2A\) gives the same result.
\item Taking the coefficient of \(A^4\) in \(\mathrm d^2B\) gives
\begin{equation}
3 h_{i[jk|n}f^n{}_{|lm]}
+
2f^n{}_{i[j}h_{klm]n}
= 0.
\end{equation}
\end{itemize}
\end{example}

\begin{example}[Six dimensions]
Six is the lowest dimension in which a nontrivial AKSZ-type theory without scalars or 1-forms exists. Avoiding scalars and 1-forms, the most general action contains a 2-form \(B\) and a 3-form \(C\), with the action
\begin{align}
S =  \int C_i\wedge \mathrm dB^i + \frac16 \alpha_{ijk} B^i \wedge B^j \wedge B^k + \frac12 \eta^{ij}C_i \wedge C_j.
\end{align}
The equations of motion are
\begin{align}\label{6d-eom}
\mathrm dB^i + \eta^{ij}C_j &= 0 &
\mathrm dC_i + \frac12\alpha_{ijk}B^j \wedge B^k &= 0.
\end{align}
In order for the Bianchi identities \(\mathrm d^2B^i = 0\) to hold, we must have
\begin{equation}\label{6d-restriction}
\eta^{ij}\alpha_{jkl} = 0.
\end{equation}
So, the fields can be partitioned into two noninteracting groups so that in each group at least one of the coefficients \(\eta^{ij}\) and \(\alpha_{ijk}\) vanish. When \(\alpha_{ijk}\) vanishes the theory is quadratic and is trivial. Thus, we can set \(\eta^{ij}=0\) without loss of generality.
\end{example}

\begin{example}[\(BF\) models]
Consider an \(n\)-dimensional worldvolume \(\Sigma\) with just a \(k\)-form \(A^i\) and an \(n-1-k\)-form field \(B_i\), with trivial potential:
\[
S = k\int B_i\wedge\mathrm dA^i.
\]
The equations of motion are
\begin{align}
\mathrm dB_i & = 0 &
\mathrm dA^i & = 0.
\end{align}
This is seen to be a generalized Abelian BF model.

If \(k=1\), then we may write a potential term
\[
S=k\int (B_i \mathrm dA + f^i{}_{jk}B_iA^j\wedge A^k).
\]
This is the non-Abelian BF model.
\end{example}

\section{Review of mathematical preliminaries}
We define certain mathematical terms that we will need for our exposition. All of our ground fields will be real, and all of our associative algebras will be unital; all of our graded algebras will be degreewise finite, and similarly the graded algebroids will be of finite type. The symbol \(\mathbb N=\{0,1,2,\dotsc,\}\) as well as the term \emph{natural number} will include zero. The mathematically sophisticated reader should feel free to skip this section.

\subsection{Algebras with differential}
The material in this section goes by the name of \emph{rational homotopy theory}, and is developed in more detail in e.g.~\cite{hess,fht,bt,fot}.

\begin{defn}
A \textbf{graded-commutive differential graded algebra}, or \textbf{cdg-algebra} for short, is a vector space \(A\), with nonnegative-integer grading
\[
A = \bigoplus_{i\in\mathbb N}A_i
\]
with an associative product that respects the grading:
\[
\deg (ab) = \deg a+\deg b
\]
for homogeneous \(a,b\in A\), equipped with a differential
\[\mathrm d\colon A_i\to A_{i+1},\]
such that
\[\mathrm d(xy) = (\mathrm dx)y + (-)^{\deg x}x\mathrm dy\]
and
\[
xy = (-)^{\deg x\deg y}yx.
\]
The algebra \(\Omega(M)\) of differential forms forms a cdg-algebra.

A cdg-algebra is called \textbf{semifree} iff it is free as a graded-commutative algebra. A semifree cdg-algebra is called \textbf{Sullivan} iff the generators can be ordered such that the derivative of each generator only depends on the ones preceding it (and excluding itself). A Sullivan cdg-algebra is called \textbf{minimal Sullivan} iff this ordering can be chosen to be nondecreasing with respect to the degree.
\end{defn}
This definition, while arbitrary-looking, can be derived from the general abstract framework known as \emph{model-category theory}, which, given certain input data, generates a definition of “nice things” (\emph{bifibrant objects}) and a canonical way of constructing approximations to things (\emph{(co-)fibrant resolutions}) in terms of these nice things, so that homotopy theory can work properly. For the originally motivating example of topological spaces, the “nice” objects in question are cell complexes, constructed by adding “cells” iteratively; for cdg-algebras the “nice” objects are Sullivan algebras, and they can be viewed as analogues of cell complexes, obtained by adding generators iteratively.

Now, every semifree cdg-algebra that lacks degree-0 and degree-1 generators is Sullivan; and if in addition every derivative lacks products of length one (contains only terms of length two or longer), then it is minimal Sullivan. Therefore the difference between semifreeness and Sullivanness is a degree-1 affair.\footnote{In general, rational homotopy theory functions best when ignoring degree~1, that is, for simply connected things. This fact is, of course, well known in topology: only the degree-1 homotopy group (fundamental group) can be non-Abelian, for example.}

By model-category theory, for every cdg-algebra \(A\), there exists a Sullivan algebra \(\tilde A\) and a cdg-algebra homomorphism \(i\colon\tilde A\to A\) that induces isomorphisms on their cohomology algebras (“is a quasi-isomorphism”) and is surjective. But this is in general not unique. For every cdg-algebra, there exists a \emph{unique} minimal Sullivan algebra \(\tilde A\) and a unique cdg-algebra homomorphism \(\tilde A\to A\) that is a quasi-isomorphism; but this is not in general surjective.

The \emph{raison d’être} of this algebraic framework is to classify topological spaces up to a rough sort of equivalence (“rational homotopy equivalence”), one that induces an equivalence on all homotopy groups tensored with a characteristic-0 field \(K\) in question (for us, the real numbers). On every manifold \(M\), the algebra of differential forms \(\Omega(M)\) is a real cdg-algebra. More generally, on every topological space in which a simplicial structure can be given, cdg-algebras of differential forms can be defined. The central statement of rational homotopy theory is that
\begin{quote}
(real) minimal Sullivan algebras are bijection with equivalence classes under real homotopy equivalence.
\end{quote}
That is, a minimal Sullivan algebra uniquely and canonically encodes the data of this equivalence class.

\subsection{Generalizations of Lie algebras}
An \(L_\infty\)-algebra is a homotopy-theoretic generalization of a graded Lie superalgebra. That is, it comes equipped with a differential, and the super-Jacobi identities hold up to exact terms. That is, the (binary) Lie bracket satisfies the super-Jacobi identity up to the differential of a certain ternary bracket; this ternary bracket satisfies a generalized Jacobi identity up to the differential of a 4-ary bracket, and so on. In fact, the differential fits into this scheme as the unary bracket, and its super-Jacobi identity is just the familiar Leibniz rule for derivations. The brackets are all strictly and totally graded-anticommutative.

While clear in concept, written out, the definition becomes quite complicated.
\begin{defn}
An \(L_\infty\)-algebra consists of an \(\mathbb N\)-graded vector space \(V=V_0\oplus V_1\oplus V_2\oplus\dotsb\) equipped with a series of \(n\)-linear operators \([-,-,\dotsc,-]\) for each positive integer \(n\in\mathbb Z^+\), each of degree \(2-n\), that is,
\[
\deg [x_1,x_2,\dotsc,x_n] = (2-n) + \deg x_1+\deg x_2+\dotsb+\deg x_n,
\]
that is
\begin{itemize}
\item totally graded-anticommutative, that is, switching any two elements of degrees \(k\) and \(l\) produces a sign \((-)^{1+kl}\);
\item 
such that the \textbf{homotopy Jacobi identity} holds:
\[
0 = \sum_{i+j=n+1}\sum_{\sigma\in\operatorname{Sh}(i,n-i)} (-)^{\sigma+i(j-1)}
[[x_{\sigma(1)},\dotsc,x_{\sigma(i)}],x_{\sigma(i+1)},\dotsc,x_{\sigma(n)}]
\]
where \(\operatorname{Sh}(i,n-i)\) spans over the so-called \emph{shuffle permutations}, which essentially means one does not overcount terms that are same except for some trivial permutation of the arguments of the brackets; and \((-)^\sigma\) is the usual sign of a permutation, with an additional minus sign for each exchange of pair of odd elements.
\end{itemize}
This definition has the following special cases:
\begin{itemize}
\item A \textbf{differential graded Lie algebra} is a graded Lie superalgebra equipped with a differential of degree \(+1\) that satisfies the graded Leibniz rule with respect to the Lie bracket; it is straightforward to check that this is exactly the same as an \(L_\infty\)-algebra, where all brackets of arity higher than three vanish.
\item If all brackets other than the binary ones vanish (including the unary one), then an \(L_\infty\)-algebra reduces to a graded Lie superalgebra. If, additionally, all elements are in degree \(0\), then this is equivalent to a Lie algebra.
\item If all brackets other than the unary one vanish, then an \(L_\infty\)-algebra reduces to a cochain complex.
\end{itemize}
\end{defn}
The above definition, while conceptually simple, is totally unmanageable. Thankfully, there exists an alternate, equivalent, and much simpler definition:
\begin{quote}
An \(L_\infty\)-algebra is the same thing as a semifree cdg-algebra.
\end{quote}
This equivalence goes by the name of \textbf{Koszul duality}. Concretely, given an \(L_\infty\)-algebra \(\mathfrak g\) with homogeneous basis \(t^i\), we construct a semifree cdg-algebra as follows. For each \(L_\infty\)-basis element \(t_i\), we put in a cdg-algebra generator\footnote{We use raised indices because the Chevalley–Eilenberg algebra is (as an algebra) freely generated by the \emph{dual space} of the \(L_\infty\)-algebra. The canonical inner product between a vector space and its dual space appears in the differential, for example. We can choose not to dualize, but then we would be working with coalgebras, not algebras.} \(t^i\) of degree \(\deg t^i = \deg t_i + 1\).\footnote{This annoying convention is needed to make the usual graded-commutativity rules work.} These are equipped with a differential
\[
\mathrm d_{\CE}t^i = - \sum_{n=0}^\infty \frac1{n!} \sum_{\deg t^i + 1 = \deg t^{j_1}+\dotsb+\deg t^{j_n}}\langle t^i|[t_{j_1},\dotsc,t_{j_n}]\rangle t^{j_1} \wedge t^{j_2} \wedge \dotsb \wedge t^{j_n}.
\]
where \(\langle-|-\rangle\) is the canonical pairing between the cdg-algebra generators and the \(L_\infty\)-basis elements. This semifree cdg-algebra is called the \textbf{Chevalley–Eilenberg algebra} and denoted \(\CE(\mathfrak g)\).

It is a tedious but straightforward exercise in combinatorics to check that the nilpotence \(\mathrm d_{\CE}^2 = 0\) of the Chevalley–Eilenberg differential is exactly equivalent to the homotopy Jacobi identities. Conversely, given a semifree cdg-algebra, one can decompose the differentials by number of terms to reconstruct the \(L_\infty\)-algebra.

It is also worth noting that the case of Koszul duality for a Lie algebra is already familiar: it is the duality between the Lie algebra and the left-invariant differential forms on its associated Lie group. In particular, if the Lie group is compact, then (by an averaging argument) the cohomology of the Chevalley–Eilenberg algebra gives the topological cohomology of the Lie group.

All this is the infinitesimal part of a theory of \(\infty\)-Lie groups, the precise formulation of which is best attempted after a long digression in higher category theory, and will not be attempted here.

\subsection{Oidification}
We shall also occasionally use the terminology \emph{\(X\)-oid} instead of \(X\). Essentially, this merely means that we consider a family of \(X\), parametrized by points of a smooth manifold (the moduli space).\footnote{The definition of such a concept, here informally called the \emph{oidification}, is also more formally known as \emph{horizontal categorification}.} The paradigmatic example is the relation between a vector space and a vector bundle. Similarly, a \emph{Lie algebroid} is a moduli space of Lie algebras. A Lie algebra is given by a vector space with additional structure; similarly a Lie algebroid is given by a vector bundle with additional structure. Exactly what this structure is can be given in elementary differential-geometric terms (see e.g.~\cite{weinstein96} and references therein), but for us it is simpler to remark that this complication is already subsumed by that of Koszul duality: the concept of a Lie algebroid precisely coincides with that of a cdg-algebra, whose degree-0 part is of the form \(\mathcal C^\infty(M;\mathbb R)\) on a smooth manifold \(M\) (the moduli space), and which is otherwise freely generated by only degree-1 generators. Relaxing the last condition gives us the concept of \(L_\infty\)-algebroid, which is (the Koszul dual of) a cdg-algebra, freely generated except in degree~0, and whose degree-0 part is of the form \(\mathcal C^\infty(M,\mathbb R)\). Once again, this boils down to a graded vector bundle on a manifold with a set of \(n\)-ary operations on its sections satisfying certain complicated identities.

We will finally note that all this is to be regarded as the infinitesimal part of a theory of Lie (or \(L_\infty\)) groupoids. A Lie groupoid (again see \cite{weinstein96}) is a parametrized family of Lie groups, except that, by virtue of the exponentiation, it consists of elements that “start” and “end” at two points on the moduli space; if one only remembers those elements that start and end at the same point, then one has a smooth family of Lie groups. 

\subsection{Invariant forms}\label{invariant-form}
Given the Chevalley–Eilenberg algebra \(\CE(\mathfrak g)\) of an \(L_\infty\)-algebra \(\mathfrak g\), the \textbf{Weil algebra} \((\Weil(\mathfrak g),\mathrm d_{\Weil})\) is the cdg-algebra obtained by adjoining, to each generator \(t_i\), a new generator \(\mathrm Dt_i\) with degree \(\deg \mathrm D t_i = \deg t_i+1\), equipped with a new differential \(\mathrm d_{\Weil}\) defined as
\begin{align}
\mathrm d_{\Weil}(\delta t_i) &= - \delta (\mathrm d_{\CE}t_i) & 
\mathrm d_{\Weil}t_i &= \mathrm dt_i + \delta t_i.
\end{align}
These may be summarized compactly as
\begin{align}
\mathrm D \,\mathrm d_{\CE}  &= - \mathrm d_{\CE} \, \mathrm D & 
\mathrm D^2 &= 0 &
\mathrm d_{\Weil} &= \mathrm d_{\CE}+\mathrm D.
\end{align}
(Structures of this kind, involving (partially-)commuting sets of nilpotent differentials, are analysed in detail in \cite{ks}.)

The Weil algebra comes equipped with the obvious quotient cdg-algebra homomorphism
\begin{equation}
\Weil(\mathfrak g) \to \CE(\mathfrak g)
\end{equation}
obtained by killing the generators of the form \(\mathrm Dt^i\).
The \emph{raison d’être} of the Weil algebra is that it has trivial cohomology; in fact, if \(\mathfrak g\) is the Lie algebra of a compact Lie group \(G\), the above quotient homomorphism is an algebraic model of the (total space of the) classifying principal bundle \(\mathrm EG\hookleftarrow G\).\footnote{This observation, in the form of Lie algebra cohomology, is how the classical names of Chevalley, Eilenberg, and Weil became attached to these algebras.}

In the Weil algebra \((\Weil(\mathfrak g),\mathrm D)\), consider an element \(\omega \in \Weil(\mathfrak g)\) of degree \(n\) that
\begin{itemize}
\item consists entirely of (sums of products of) generators of the form \(\mathrm Dt^i\), and
\item is closed (i.e. \(\mathrm d_{\Weil}\omega = 0\)).
\end{itemize}
In the Lie algebra case, this definition reduces to the classical notion of an invariant polynomial of degree \(n/2\) (here \(n\) is necessarily even); e.g.~the Killing form \(B(-,-)\) is encoded as the degree-4 element \(B_{ij}\mathrm Dt^i\mathrm Dt^j \in \operatorname{Weil}(\mathfrak g)\).

\section{Dictionary between Rational Homotopy Theory and Topological Quantum Field Theory}
Using the heaviweight mathematical technology which we have sketched in the previous section, we now proceed to interpret the structure of AKSZ-type topologial field theories in terms of algebraic and topological structures, explaining and elucidating various features of this class of theories.

\subsection{Fields as homomorphisms/continuous maps}
Given a manifold \(M\) and an \(L_\infty\)-algebra \(\mathfrak g\) with homogeneous basis \((t^i)_{i\in I}\), consider a cdg-algebroid morphism
\[\Phi\colon \CE(\mathfrak g) \to \Omega(M).\]
Because the domain is a semifree cdg-algebra, this map is specified by the images of generators
\[\Phi(t^i) \in \Omega^{\deg t^i}(M).\]
That is, we obtain a series of differential forms \(A^i\) of degree \(\deg t^i\), that must satisfy identities of the form
\begin{equation} \label{ce-morphism-form}
\mathrm dA^i = a_jA^j + b_{jk} A^j\wedge A^k + \dotsb
\end{equation}
corresponding to the preservation of \(\mathrm d_{\CE}\).
This can be thought of as flatness conditions for a field strength
\[F^i = \mathrm dA^i - a_jA^j - \dotsb.\]
In fact, this assertion can be suitably formalized; this datum defines a flat connection in the trivial \(\mathfrak g\)-valued principal bundle on \(M\).\footnote{
On nontrivial \(\mathfrak g\)-valued principal bundles — sometimes called \emph{gerbes} — of course potentials cannot be globally defined, and must be defined patchwise.
}

On the other hand, consider a morphism
\[\Phi \colon \Weil(\mathfrak g) \to \Omega(M).\]
Let the images of \(t^i\) be \(A^i\) and the images of \(\mathrm Dt^i\) be \(F^i\).
Then the equivalent identities become
\begin{equation}
\mathrm dA^i = F^i + a_jA^j + b_{jk} A^j\wedge A^k + \dotsb
\end{equation}
corresponding to the preservation of \(\mathrm d_{\Weil} = \mathrm d_{\CE} + \mathrm D\). Now the field strength \(F^i\) can be nonzero, but it must nevertheless satisfy the Bianchi identity \(\mathrm DF = 0\), corresponding to the nilpotence of \(\mathrm D\) in \(\Weil(\mathfrak g)\).

So we obtain the following dictionary:
\begin{center}
\begin{tabular}{cc}\toprule
cdg-Algebra & Physics \\ \midrule
operator \(\mathrm D\) & covariant derivative \(\mathrm D\) \\
\(\Weil(\mathfrak g)\)-differential \(\mathrm d_{\Weil}\) & exterior derivative \(\mathrm d\) \\
\(\CE(\mathfrak g)\)-differential \(-\mathrm d_{\CE} = \mathrm D -  d_{\Weil}\) & nonderiv.\ part of covariant deriv. \\
nilpotence of \(\mathrm D\) & Bianchi identity \(\mathrm DF = 0\)\\
cdg-algebra morphism \(\CE(\mathfrak g)\to\Omega(M)\) & potential with \(0\) field strength \\
cdg-algebra morphism \(\Weil(\mathfrak g)\to\Omega(M)\) & potential \\
\bottomrule
\end{tabular}
\end{center}
In particular, if one specializes to the case where \(\mathfrak g\) is a Lie algebra, one obtains the usual notions for Yang–Mills theory.

\subsection{AKSZ-type Lagrangians as symplectic structures}
If one examines the equations of motion for the AKSZ-type topological quantum field theories, one notices that the equations of motion are of the form \eqref{ce-morphism-form}. That is, their structure can be encoded by an \(L_\infty\)-algebra \(\mathfrak g\), and can be regarded as a theory of flat connections for a \(\mathfrak g\)-bundle.

We may ask the converse question: which cdg-algebras correspond to encodings of AKSZ-type theories? This is easily answered. First, the kinetic term of the AKSZ-type Lagrangian
\[
\mathcal L = C_{ij}A^i \wedge \mathrm dA^j + \dotsb 
\]
defines a bilinear form \(C_{ij}\) on the \(L_\infty\)-algebra \(\mathfrak g\). Furthermore, by integration by parts, \(C_{ij}\) can be assumed without loss of generality to be graded-antisymmetric with respect to the \(L_\infty\)-algebra grading (and not the \(\CE(\mathfrak g)\)-grading).

Such forms can be encoded as closed elements in the Weil algebra
\[C = C_{ij}\mathrm DA^i \mathrm DA^j \in \Weil(\mathfrak g)\]
that consists of elements of the form \(\mathrm Dt^i\), as explained in section~\ref{invariant-form}; they necessarily have degree \(n+1\), where \(n\) is the dimension of the worldvolume. \(L_\infty\)-algebras equipped with the choice of such a structure has been called \emph{symplectic \(L_\infty\)-algebras} or \(\Sigma_n\)-manifold in the literature \cite{frs,severa05}: if one considers the equivalent \(L_\infty\)-algebroid notion, then the \(n=1\) \(L_\infty\)-algebroid corresponds to a symplectic manifold, and the \(n=2\) \(L_\infty\)-algebroid to a Poisson manifold. The \(n=3\) \(L_\infty\)-algebra is a \emph{quadratic Lie algebra}, that is, a Lie algebra equipped with an invariant symmetric binary form. The \(n=3\) \(L_\infty\)-algebroid is called a \emph{Courant algebroid} in the literature \cite{roytenberg,ikeda}.

What about the potential \(p\in\CE(\mathfrak g)\) in the AKSZ-type Lagrangian? It can be straigthforwardly verified, once one unwinds the definition, that the existence of the potential amounts to the existence of an antiderivative \(\mathrm dc=C\) of the symplectic element \(C\in\Weil(\mathfrak g)\), such that \(c\) maps to \(p\) under the canonical forgetful cdg-algebra homomorphism \(\CE(\mathfrak g)\to\Weil(\mathfrak g)\). An antiderivative of \(C\) always exists, since the Weil algebra has trivial cohomology by construction. 

That is, we have the following correspondence \cite{severa00, severa05, frs}:
\begin{center}
\begin{tabular}{cc}\toprule
Geometry & Physics \\ \midrule
symplectic manifold & 1-dimensional AKSZ-type theory \\
Poisson manifold & 2-dimensional AKSZ-type theory \\
Courant algebroid & 3-dimensional AKSZ-type theory \\
quadratic Lie algebra & 3D AKSZ-type theory without scalars \\
\bottomrule
\end{tabular}
\end{center}

\subsection{Gauge transformations as Koszul duality}
Given that AKSZ-type theories are a theory of flat connections over gerbes with \(L_\infty\)-algebra fibre, it should therefore be the case that the gauge transformations are also valued in \(L_\infty\)-algebras (at least for topologically trivial gerbes), the same way that a gauge transformation for a \(G\)-bundle for a Lie group \(G\) are smooth \(G\)-valued functions (if the principal bundle is topologically trivial and/or \(G\) is Abelian).

We explain how this is in fact the case. The relation between gauge transformations and gauge fields is the same one as that between \(L_\infty\)-algebras and semifree cdg-algebras, or Koszul duality, as detailed in the following examples.

In general, in an \(L_\infty\)-algebra, the Jacobi identity only holds up to homotopy (“on-shell”). However, the semifree cdg-algebra is minimal Sullivan, that is, if the derivatives of each generator does not have linear terms (only quadratic or higher), then the differential (unary bracket) in the \(L_\infty\)-algebra is zero, and the Jacobi identity for the binary bracket holds exactly. In this case, the Jacobi identity holds exactly (“off-shell”), and the algebra of gauge transformations forms a Lie superalgebra (or a Lie supergroup, when exponentiated).

Nevertheless, even in this case, the algebra of gauge transformations is more than a mere Lie superalgebra, because the higher-arity brackets do not vanish in general.\footnote{The following two statements are equivalent: (1) The \(L_\infty\) \(n\)-ary bracket (i.e.~differential) vanishes; (2) the differential in the associated cdg-algebra lacks \(n\)th-order terms.} In other words, the algebra of gauge transformations has additional structure beyond the Lie (super-)bracket.

\begin{example}[3D, continued]
The equations of motion of Chern–Simons theory, namely \(F=0\), determine a cdg-algebra, which is \(\CE(\mathfrak g)\). Its dual \(L_\infty\)-algebra is the Lie algebra \(\mathfrak g\). Gauge transformations (for a trivial principal bundle) are parametrized by maps \(\Sigma\to G\); infinitesimally they are given by maps \(M\to\mathfrak g\).\footnote{Of course, in general, on a principal bundle \(P\) a gauge transformation is given by a section of the associated bundle \(\operatorname{Ad}(P) = P \times_G G\) (where \(G\) acts on itself by conjugation), and an infinitesimal gauge transformation is given by a section of the associated bundle \(\mathfrak{ad}(P)=P\times_G\mathfrak g\). When \(P\) is trivial this reduces to \(\operatorname{Ad}(P)=M\times G\) and \(\mathfrak{ad}(P)=M\times\mathfrak g\); this identification is canonical, independent of the choice of a global section of \(P\).
}

The Yang–Mills potentials transform under a gauge transformation
\[
\alpha \colon x \mapsto \alpha^i(x)t_i \in\mathfrak g
\]
as
\[
A^i \mapsto A^i + \mathrm d\alpha^i.
\]
That is, the algebra of infinitesimal gauge symmetries is the \(L_\infty\)-algebra dual to the cdg-algebra of the fields; it exponentiates into a Lie group \(G\). Note that the torsion/fundamental group of \(G\) is not determined by the cdg-algebra, which only determines the (torsion-free) cohomology with real coefficients.
\end{example}
\begin{example}[2D, continued]\label{2d-gauge-discussion}
We continue the analysis of the Poisson sigma model \eqref{2d-action}.
This theory has the following infinitesimal gauge symmetry \cite[(2.10)]{ikeda}:
\begin{align}
x^i & \mapsto x^i - \pi^{ij}(x)t_j \\
e_i & \mapsto e_i + \mathrm dt_i + \frac12 (\partial_i\pi^{jk})t_k,
\end{align}
where the gauge parameter \(t\) is a section of the vector bundle
\[t \in \Gamma(x^*\mathrm T^*M).\]
The allowed values of the gauge parameter depends on \(x\in M\); that is the gauge symmetry is described by a Lie algebroid, rather than a Lie algebra.

Taking the commutators of the infinitesimal gauge symmetries,
\begin{align}
\delta_t\delta_u x^i &= - \pi^{ij}t_j - \pi^{ij}u_j 
+ (\partial_k\pi^{ij})\pi^{kl}t_ju_l
\end{align}
we obtain
\[
[\delta_t,\delta_u]x^i = 
\left(
(\partial_k\pi^{ij})\pi^{kl}
-
(\partial_k\pi^{il})\pi^{kj}
\right)
t_ju_l
=(\pi^{ik}\partial_k\pi^{jl})t_ju_l.\]
In other words, the structure constants of the gauge Lie algebra are given by \(\partial_k\pi^{ij}\), which mirror those in the equation of motion for \(e\), as required by Koszul duality.
\end{example}

\begin{example}[4D, continued]
Consider the action \eqref{4d-action}, with \(\eta=0\). It admits the gauge transformation
\begin{align}
\delta A_i &= \mathrm d\alpha^i - f^i{}_{jk}\alpha^jA^k &
\delta B^i &= \mathrm d\beta^i - f^i{}_{jk}\alpha^jA^k.
\end{align}
Now, instead of writing out the components \(A^i\) and \(B^i\), we take the point of view that \((A,B)\) defines an \(L_\infty\)-algebra-valued form (a gerbe connection), where the \(L_\infty\)-algebra in question is the one Koszul-dual to the cdg-algebra defined by the field equations. Let the basis elements of the \(L_\infty\)-algebra be \(\mathsf a^i\) and \(\mathsf b_i\).\footnote{These are analogues of the Pauli matrices for \(\mathfrak{su}(2)\) or the Gell-Mann matrices for \(\mathfrak{su}(3)\).} Then the gauge transformations are of the form
\begin{align}
\delta A &= \mathrm d\alpha^i\mathsf a_i - f^i{}_{jk}\alpha^jA^k\mathsf a_i &
\delta B &= \mathrm d\beta^i\mathsf b_i - f^i{}_{jk}\alpha^jA^k\mathsf b_i.
\end{align}
The gauge algebra is then expressed in terms of the brackets \([\alpha_i,\alpha_j]\), etc. Now, the gauge algebra can be derived instead directly from the equations of motion: the condition that a Yang–Mills field be flat translates to the semifree cdg-algebra defined by
\begin{align}
\mathrm d\mathsf a^i & = - \frac12 f^i{}_{jk} \mathsf a^j \mathsf a^k &
\mathrm d\mathsf b_i & = - f^k{}_{ij} \mathsf a^j \mathsf b_k - \frac16 h_{ijkl} \mathsf a^j \mathsf a^k \mathsf a^l
\end{align}
where \(h_{ijkl}\) is totally symmetric and \(f^i{}_{jk}\) are structure constants for a Lie algebra. (This case is not Sullivan due to degree-1 generators.) 

This encodes (is Koszul-dual to) the \(L_\infty\)-algebra of the gauge transformations:
\begin{align}
\deg \alpha_i &= 0 & \deg \beta^i &= 1 \\
[\alpha_j,\alpha_k] &= f^i{}_{jk} \alpha_i &
[\alpha_j,\beta^k] &= f^k{}_{ij} \beta^i &
[\alpha_j,\alpha_k,\alpha_l] &= h_{ijkl} \beta^i.
\end{align}
All other brackets vanish. The binary brackets are an infinitesimal version of the finite gauge-symmetry composition law
\begin{align}
\exp(s^i\alpha_i)
\exp(s'^i\alpha_i)
&= 
\exp\left((s+s')^i\alpha_i+ \frac12s^j s^k f^i{}_{jk}\alpha_i+\dotsb\right) \\
\exp(t_i\beta^i)
\exp(t'_i\beta_i)
&=\exp((t+t')_i\beta^i) \\
\exp(s^i\alpha_i)
\exp(t_i\beta^i)
\exp(-s^i\alpha_i)
&=\exp\left(
s^j t_k f^k{}_{ij}\beta^i
+O(s^2) + O(t^2)
\right)
\end{align}
where \(s\) and \(s'\) are commuting parameters and \(t\) and \(t'\) are anticommuting. However, this forgets the ternary bracket. The gauge symmetry algebra is \emph{more than just a Lie superalgebra}; it carries an additional structure, the nonvanishing ternary bracket. In other words, this is an \(L_3\)-algebra.

The ternary Jacobi identity means that \(f^i{}_{jk}\) satisfies the usual Jacobi identity. The next nontrivial Jacobi identity is the quaternary one, which in our case is
\begin{equation}\label{cocycle-infinitesimal}
[[x,y,z],w] - [[y,z,w],x] + [[z,w,x],y] - [[w,x,y],z] = 0
\qquad\forall x,y,z,w\in \operatorname{Span} \{\mathsf a^i\}
\end{equation}
(All others are trivial.) In indices,
\begin{equation}
h_{nijk}f^n{}_{lm}
-
h_{njkl}f^n{}_{im}
+
h_{nkli}f^n{}_{jm}
-
h_{nlij}f^n{}_{km}
 = 0.
\end{equation}
\end{example}

\begin{example}[6D continued]
Suppose that we have the semifree cdg-algebra defined by
\begin{align}
\deg\mathsf b^i & = 2 & \mathrm d\mathsf b^i & = 0 &
\deg\mathsf c_i & = 3 & \mathrm d\mathsf c_i & = - \frac12 \alpha_{ijk} \mathsf b^j \mathsf b^k.
\end{align}
This encodes the \(L_\infty\)-algebra
\begin{align}
\deg \beta_i & = 1 & \deg \gamma^i &= 2 &
\{\beta_j,\beta_k\} &= \alpha_{ijk}\gamma^i 
\end{align}
with all other brackets vanishing. Because only the binary bracket is nonzero, this is a graded superalgebra. The bosonic subalgebra, spanned by \(\gamma^i\), exponentiates into an Abelian Lie group.

This is the infinitesimal version of the (finite) supergroup composition law
\begin{equation}
\exp(s^i\beta_i+t_j\gamma^j)  \exp(s'{}^i\beta_i+t'_j{}\gamma^j)
= \exp\left(
(s+s')^i\beta_i
+(t+t')_j\gamma^j +
\frac12s^js'{}^k\alpha_{ijk}\gamma^i
\right)
\end{equation}
where \(s\) and \(s'\) are anticommuting parameters and \(t\) and \(t'\) are commuting.

There being only binary products, the Jacobi identity is the same as that for a Lie superalgebra, and these are all trivial (apart from \(\alpha_{[ij]k}=0\).). Note that the requirement that \(\alpha\) be totally symmetric does not come from Jacobi identities; it comes from the existence of a realization as an AKSZ-type TQFT.
\end{example}

\subsection{Sullivan resolution}
We have seen how an AKSZ-type theory encodes an \(L_\infty\)-algebra equipped with a Chern–Simons element for a symplectic structure, that is, to semifree cdg-algebras, and, if the coefficients can be made to be rational, this in turn encodes a topological space (up to rational homotopy), i.e.~a rational homotopy type. However, in the latter step, multiple cdg-algebras may correspond to the same rational homotopy type, even if we ignore algebra isomorphisms. However, there exists a canonical such algebra, the so-called Sullivan minimal model. We discuss the physics interpreation of this canoncalization.

The process of taking this canonical form has two parts: one easy, one hard.
\begin{itemize}
\item The easier part is in eliminating pairs of fields, of the form \(\mathrm dX = Y\) and \(\mathrm dY=0\), that can be eliminated. (Topologically, this can be thought of as adding a filled ball to a cell complex, which is topologically trivial.) Elimination of such pairs correspond to integrating out auxiliary fields via substitution into the action.
\item The more complicated part consists of dealing with degree-1 generators. In topological literature one usually makes the simplifying assumption that they are absent. In physics they correspond to Yang–Mills fields, and Sullivanization corresponds to replacing non-Abelian Yang–Mills fields by higher-degree Abelian fields corresponding to Chern–Simons forms.
\end{itemize}
We discuss each in turn.

\subsubsection{Integrating out Yang–Mills fields}
The obstruction to a semifree cdg-algebra \(\mathfrak g\) being Sullivan lies in degree~1, or equivalently Yang–Mills gauge fields. Let us suppose that, in fact, the geometrical realization of \(\operatorname{CE}(\mathfrak g)\) is simply connected.\footnote{This restriction can be slightly relaxed to allow Abelian fundamental groups that act trivially on higher homotopy groups, the so-called \emph{simple spaces}.} Then Sullivan resolution produces a theory with the same set of non-auxiliary\footnote{that is, exclusing thing like pairs of auxiliary fields found in non-minimal Sullival algebras, but including possibly composite fields (such as the Chern–Simons forms)} Abelian higher gauge fields, but with no Yang–Mills fields.

Note, however, that the resulting (minimal) Sullivan model may not admit a symplectic structure. This does not mean that they cannot be TQFTs; they can usually be embedded into a bigger minimal Sullivan algebra that can be given a symplectic structure (by an algebraic analogue of taking the cotangent bundle of a manifold).

\begin{example}
As a paradigmatic example, we consider Chern–Simons theory.

Consider the cdg-algebra \(A\) given by
\begin{align}
\mathrm x &= yz & \mathrm dy &= zx & \mathrm dz &= xy &\deg x =\deg y=\deg z = 1.
\end{align}
This is a semifree model for the 3-sphere, and is the Chevalley–Eilenberg algebra of \(\mathfrak{su}(2)\).

As a 3-sphere, its minimal model \(A_0\) is simply
\begin{align}
\mathrm dh &= 0 & \deg h &= 3.
\end{align}
The cdg-algebra homomorphism
\[h \mapsto xyz\]
is a (non-injective) quasi-isomorphism.

If we are willing to sacrifice minimality, then we can instead use the non-minimal Sullivan algebra
\begin{gather}
\begin{aligned}
\mathrm dx &= X & \mathrm dy &= Y & \mathrm dz &= Z 
\end{aligned}
\deg x =\deg y=\deg z = 1 \\
\begin{aligned}
\mathrm dX = \mathrm dY = \mathrm dZ &= 0 &
\deg X = \deg Y=\deg Z &= 2 
\end{aligned} \\
\begin{aligned}
 \mathrm dh &= 0 &
 \deg h &= 3.
 \end{aligned}
\end{gather}
The cdg-algebra homomorphism
\begin{align}
x & \mapsto x & y & \mapsto y & z &\mapsto z \\
X & \mapsto yz & Y & \mapsto zx & Z &\mapsto xy &
h & \mapsto xyz
\end{align}
is a surjective quasi-ismomorphism.

The Sullivan resolution, therefore, describes a theory of Abelian 3-form gauge field \(C\). Being one-dimensional, it does not, of course, admit a symplectic structure. In a three-dimensional worldvolume, the field equation is trivial (all 3-forms are automatically closed), but one can easily consider higher-dimensional AKSZ-type theories with a 1-form coupled to other fields:
\[
S = \int \left(\mathrm dA^i + \frac12 f^i{}_{jk}A^j \wedge A^k \right) \wedge C_i + \dotsb.
\]
Then the field equation of the Abelian 3-form field thus defined becomes nontrivial.

More generally, let \(G\) be any compact Lie group of rank \(r\), with Lie algebra \(\mathfrak g\). Let the degrees of its invariant polynomials be \(m_1,\dotsc,m_r\). Then \(G\) is rational-homotopy-equivalent to the product of odd-dimensional spheres:\footnote{This is due to the fact that the cohomology ring of such a group is a Hopf algebra, and thus generated by odd-degree elements (Hopf’s theorem), coupled with the formality (in the rational-homotopy sense) of such groups as spaces.}
\[G \simeq_{\mathbb Q} \mathbb S^{2m_1-1} \times \mathbb S^{2m_2-1} \times \dotsb \times \mathbb S^{2m_r-1}.\]
The corresponding minimal Sullivan algebra is therefore of the form
\begin{align}
\deg A_i &= 2m_i - 1 \qquad (i\in \{1,\dotsc,r\}) \\
\mathrm dA_i & = 0.
\end{align}
That is, Sullivanization amounts to integrating out some degrees of freedom, leaving behind the composite Abelian gauge fields corresponding to Chern–Simons forms (of invariant polynomials).
\end{example}

\subsubsection{Integrating out auxiliary terms}
The previous section discussed the obstruction to semifree cdg-algebras being Sullivan; what about Sullivan cdg-algebras being minimal Sullivan? In the absence of degree-1 elements (Yang–Mills fields), this amounts to the absence of linear terms in the derivatives, which integrates to the absence of quadratic terms in the potential of an AKSZ-type field.

\begin{example}[2d, continued]
At any given point \(x_0\in M\), a neighbourhood \(U \ni x\) of the Poisson manifold \(M\) can be decomposed \cite[Theorem~2.1]{Weinstein}as
\[U = S \times \tilde M,\]
where \(S\) is a symplectic manifold, \((\tilde M,\tilde\pi)\) is a Poisson manifold with \(\tilde\pi_{x_0}=0\), and such that the Poisson structure on \(U\) coincides with the product Poisson structure. Due to this, the fields corresponding to \(S\) and to \(\tilde M\) decouple, and we may consider the two cases separately.

\subparagraph{Symplectic case}
Consider the case where the Poisson manifold \((M,\pi)\) is symplectic, with the inverse
\[\omega_{ij}\pi^{jk} = \delta_i^k.\]
In that case, the equation of motion
\[
\mathrm dx^i = \pi^{ij}(x)e_j
\]
actually represents \(e_j\) as the auxiliary field
\[e_i = \omega_{ij}(x)\mathrm dx^j.\]
Thus, it can be substituted away in the action, which leaves
\[S \propto \int_\Sigma x^*\omega.\]
This is the action of the A-model in topological string theory \cite{aksz,ikeda}.

\subparagraph{Locally rank zero case}
Consider the case where the Poisson manifold \((M,\pi)\) is such that \(\pi_{x_0} = 0\) at the “origin” \(x_0\in M\). Then the cotangent space \(\mathrm T_{x_0}^*M\) is naturally imbued with the structure of a real Lie algebra \(\mathfrak g\), whose coefficients are the first-order derivatives of \(\pi\) at \(x_0\). In this case, the Lie-Poisson structure on \(\mathfrak g^*\) provides a linear approximation of \(M\) near \(x_0\) \cite{Weinstein}. Near \(x_0\), the action becomes
\[
S/2\pi = \int_\Sigma e_i \wedge \mathrm dx^i - \frac12 \langle x^k,[e_i,e_j]\rangle + \mathcal O(x^2e^2),
\]
where the neighbourhood of \(M\) has been identified with \(\mathfrak g^*\). Now, \(e\) represents a \(\mathfrak g\)-valued 1-form on \(\Sigma\), which we will regard as a \(\mathfrak g\)-connection. Then \(x\) is a scalar field taking values in the coadjoint representation of \(\mathfrak g\). Then the theory reduces to a \(BF\) model, up to higher-order terms; the equations of motion for \(x\) and \(e\) reduce to
\begin{align}
F_e &= 0 & \mathrm D_e x & = 0,
\end{align}
up to higher-order terms: \(e\) is a flat connection and \(x\) is covariantly constant, to leading order.
\end{example}
\begin{example}[4d, continued]\label{4d-auxiliary-example}
We continue the analysis of the action \eqref{4d-action}.
Diagonalizing \(\eta\), we can partition the allowed values of the index \(i\) into two sets, one of which we will label as \(a,b,\dotsc\) and the other as \(A,B,C,\dotsc\), such that
\begin{align}
\eta^{ab} = \eta^{Ab} &= 0 
\end{align}
and such that \(\eta^{AB}\) is a nondegenerate symmetric quadratic form (but not necessarily positive-definite). We will use \(\eta^{AB}\) and its inverse to raise and lower indices \(A,B,C,\dotsc\) freely.

Using the equations of motion \eqref{4d-restriction}, we can see that
\[f_{ABc}=f_{[AB]c}\]
and
\[f_{ABC} = f_{[ABC]}.\]
Thus there are the following components:
\begin{align}
f_{ABc} &= f_{[AB]c}&
f_{ABC} &= f_{[ABC]}&
f^A{}_{bc} &= f^A{}_{[bc]} &
f^a{}_{bc} &= f^a{}_{[bc]} &
f^a{}_{Bc} &= 0 &
f^a{}_{BC} &= 0.
\end{align}
Now, from the equation of motion
\begin{equation}
\mathrm dA^A + B^A + \frac12 f^A{}_{ij}A^i\wedge A^j = 0,
\end{equation}
it follows that \(B^A\) are auxiliary fields, which can be substituted away in the action in favour of \(A^A\) and \(A^a\).

Now, we can integrate out the nondegenerate part of \(\eta\) as follows. The part of the action \eqref{4d-action} containing \(B^A\) are
\begin{equation}
S_\text{aux} = \int B_A\wedge
\left(\mathrm dA^A + \frac12 f^A{}_{ij} A^i\wedge A^j\right)
+ \frac12 B_A\wedge B^A
\end{equation}
Substituting, we find
\begin{equation}
S_\text{aux} = \int 
-\frac12
\left(\mathrm dA^A + \frac12 f^A{}_{ij} A^i\wedge A^j\right)^2
\end{equation}
where the square involves contraction of the \(A\)-index.



Formally, this is caused by the cyclic dependency in the expressions of the derivatives, which are violations of the Sullivan condition for cdg-algebras.
After integrating out \(B_A\) in terms of \(A^A\) and \(A^i\), we have the dependency graph
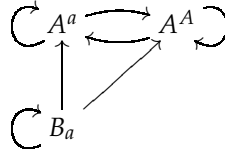
\[
\xymatrix{
A^a \ar@/^/[r]  \ar@(dl,ul) & A^A \ar@/^/[l] \ar@(ur,dr) \\
B_a\ar@(dl,ul) \ar[u] \ar[ur]
}
\]
where an arrow \(X\to Y\) means that the expression for the derivative of \(X\) contains \(Y\).
In particular, the mutual dependence between \(A^a\) and \(A^A\) makes integrating out \(A^A\) result in an non-AKSZ-type action. Due to the non-bilinear form of the kinetic term, the equations of motion will produce constraints on the fields (as opposed to their exterior derivatives); in other words, the algebra of (wedge-products of) the fields will not be quasifree.

To avoid this, we make the additional assumption that \(\mathrm dA^a\) and \(\mathrm dA^A\) both only depend on \(A^a\); that is, we set
\begin{align}
f^A{}_{Bc} \ne 0 &\ne f^A{}_{bc}& 
f^a_{Bc} = f^a{}_{BC} = f^A{}_{Bc} = f^A{}_{BC} &= 0.
\end{align}
Then \eqref{4d-jacobi} implies that
\begin{align}
h_{abcd} &\ne 0 \ne h_{Abcd} &
h_{ABcd} = h_{ABCd} = h_{ABCD} = 0.
\end{align}
Now, the dependency graph simplifies to
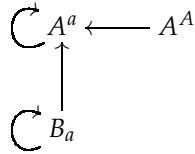
\[
\xymatrix{
A^a  \ar@(dl,ul) & A^A \ar[l] \\
B_a\ar@(dl,ul) \ar[u]
}
\]
The self-dependence of \(A^a\) and \(B_a\) are still troublesome (a characteristic feature of non-Abelian Yang–Mills fields — i.e.\ nontrivial degree-1 generators of cdg-algebras), so we still do not have a Sullivan cdg-algebra; but at least it does not affect the \(A^A\), which will be rendered auxiliary.

Now, the non-total-derivative terms in the action involving \(A^A\) are
\begin{equation}
S_\text{aux} = 
\int
A^A\wedge A^b\wedge
\left(
 f_{Abc} \mathrm dA^c
+ \frac16 h_{Abcd}A^c\wedge A^d\right).
\end{equation}
We have killed enough coefficients to make the fields \(A^A\) into mere Lagrange multipliers.

Using the homotopy Jacobi identity \eqref{4d-jacobi}, now written as
\begin{equation}
h_{Ajkl} = - 3 f_{Ai[j}f^i{}_{kl]},
\end{equation}
this becomes
\begin{equation}
S_\text{aux}' = 
\int
A^A\wedge A^b\wedge
f_{Abc}
\left(\mathrm dA^b+ \frac12 f^b{}_{cd}A^c\wedge A^d\right).
\end{equation}
The bracketed part is already the equation of motion for \(A^a\), as expected; therefore the constraint is already satisfied, and the Lagrange multiplier fields \(A^A\) can be dropped. We are left with an action of the same form as \eqref{4d-action} before, but with the \(\eta^{ij}\)-containing term absent, and with the redefinition
\[h'_{abcd} = h_{abcd} + 6 f^A{}_{ab}f_{Acd}.\]
in place of \(h_{abcd}\).

Let us rephase the preceding analysis in terms of rational homotopy theory. Originally, we had the quasifree cdg-algebra \(\mathcal A\) given by
\begin{align}
\deg \mathsf a^i&=1 &
\mathrm d\mathsf a^i &= - \eta^{ij}\mathsf b_j - \frac12 f^i{}_{jk}\mathsf a^j\mathsf a^k \\
\deg b_i &= 2 & \mathsf d\mathsf b_i &= - f^k{}_{ij}\mathsf b_k\mathsf a^j - \frac16 h_{ijkl}\mathsf a^j\mathsf a^k\mathsf a^l.
\end{align}
After integrating out the auxiliary fields, we have the quasifree cdg-algebra \(\mathcal A'\), given by 
\begin{align}
\deg\mathsf a^a&=1 &
\mathrm d\mathsf a^a &= - \frac12 f^a{}_{bc}\mathsf a^b\mathsf a^c \\
\deg b_a &= 2 & \mathsf d\mathsf b_a &= - f^c{}_{ab}\mathsf b_c\mathsf a^b - \frac16 h'_{abcd}\mathsf a^b\mathsf a^c\mathsf a^d.
\end{align}
The fact that \(\mathcal A'\) has been obtained by integrating out auxiliary fields from \(\mathcal A\) means that there exist a family of cdg-homomorphisms \(\mathcal A\to\mathcal A'\) that are quasi-isomorphisms; concretely, these are given by mapping the auxiliary fields \(B^A\) to its expression in terms of \(A^a\) and \(A^A\) and, futher, setting the Lagrange multiplier fields \(A^a\) (whose values are not constrained) to arbitrary values with the correct degree (including zero).
\end{example}
\begin{example}[6D, continued]
Consider the 6d model
\[S = \int C_i\wedge \mathrm dB^i + \frac16\alpha_{ijk}B^i\wedge B^j \wedge B^k + \frac12\eta^{ij}C_i\wedge C_j.\]
with \(\deg B=2\) and \(\deg C=3\). Recall (\eqref{6d-restriction}) that the indices \(i\) can be partitioned, say \(B^i = (B^a,B^\alpha)\), such that
\begin{align}
\eta^{ab} = \alpha_{\alpha\beta\gamma} = \eta_{a\alpha} &= \alpha_{a\beta\gamma} = \alpha_{ab\gamma} = 0.
\end{align}
The semifree cdg-algebra encoded by the equations of motion \eqref{6d-eom} is not Sullivan, due to the \(\alpha\)-indices: \(\mathrm dB^\alpha = - \eta^{\alpha\beta}C_\alpha\). On the other hand, \(\mathrm dB^a = 0\) is not problematic.

Assuming without loss of generality that \(\eta_{\alpha\beta}\) is nondegenerate and using it to raise and lower indices, since \(\mathrm dB^\alpha = - C^\alpha\) and \(\mathrm dC^\alpha = 0\), we can simply remove the pairs \((B^\alpha,C^\alpha)\) by substituting into the action:
\[
S \sim dB^\alpha \wedge dB_\alpha
\]
This is now quadratic, and we can integrate out the fields \(B_\alpha\), leaving a theory encoding a Sullivan cdg-algebra.
\end{example}

Minimal model resolutions can have problems if there are degree~1 (or degree~0) fields.

\begin{example}
Consider the 4d BF model with a cosmological constant:
\[S = \int B\wedge F + \frac12\lambda B\wedge B.\]
Then, the equation of motion is
\begin{align}
\mathrm dA &= \lambda B - [A,A] \\
\mathrm dB &= - A\wedge B.
\end{align}
In this case, while \(\mathrm dA\) has a linear term, \(\mathrm dB\) also depends on \(A\) (as well as \(B\)), and this mutual dependence foils a naive attempt to integrate it away. In fact, this term makes a big difference in the quantum theory; see \cite{baez} for a discussion.
\end{example}

\subsubsection{Unconstraining constrained theories}
A Sullivan minimal model can also be defined for non-quasifree cdg-algebras (that is, those not given by Koszul-dualizing an \(L_\infty\)-algebra). We can also interpret this in physics terms. A non-quasifree cdg-algebra has algebraic relations (syzygies) between terms. Seen as a quotient of a quasifree cdg-algebra, the syzygies correspond to certain constraints that one may impose upon the fields (that products of certain fields equal others), not reflected in the Lagrangian.

Rational homotopy theory shows that in general such constrained AKSZ-type theories can be mechanically transformed or “completed” into an unconstrained (semifree) theory admitting a morphism into the constrained theory; this can be thought of as a kind of ultraviolet completion.

\begin{example}
An \(n\)-sphere has a single cohomology class; thus it\footnote{Or, more precisely, its cohomology; but spheres are formal, so it does not matter.} corresponds to the cdg-algebra with a single generator \(\mathsf a\) with
\begin{align}
\deg \mathsf a&=n&
\mathrm d\mathsf a&=0 &
\mathsf a^2 &= 0.
\end{align}
The latter equation is trivial if \(n\) is odd (so the cdg-algebra is quasifree), but not if \(n\) is even (so the cdg-algebra is not quasifree).

Embedded in a physical theory, this will correspond to an \(n\)-form gauge field, required to be closed by the equation of motion, and in addition with the constraint that \(A\wedge A = 0\).

To take the Sullivan minimal model in the case where \(n\) is even, we add an extra generator \(\mathsf b\) to kill the cohomology class that would correspond to \(\mathsf a^2\), as thus:
\begin{align}
\deg \mathsf a&=n & \deg \mathsf b &= 2n-1 &
\mathrm d\mathsf a&= 0 &
\mathrm d\mathsf b&=\mathsf a^2.
\end{align}
That is, the syzygy \(\mathsf a^2 =0 \) has been resolved. In physics, this corresponds to another field \(B\), which can be integrated out, but at the cost of producing a constraint on \(A\).

For a more detailed example, see Example~\ref{4d-auxiliary-example}, where integrating out certain fields produces syzygies among fields.
\end{example}

\subsection{Consequences of geometric realizations}
In addition to constructing a canonical algebraic model of a rational homotopy type, rational homotopy theory provides statements about recovering the topological space corresponding to a given algebra. In particular, conditions are known for realizability as a compact Riemannian manifold \cite{fot}.

In some cases, the \(L_\infty\)-algebra is realizable as a smooth compact manifold \(G\); see \cite[Theorem 3.2]{fot} for a characterization. This means that there exists a cdg-algebra morphism
\[\CE(\mathfrak g) \to \Omega(G)\]
that is a quasi-isomorphism.\footnote{In the classical case where \(\mathfrak g\) is the Lie algebra of a compact Lie group, this map is simply the map whose image consists of pullback-invariant forms on \(G\) that are thus entirely determined by the value at the group identity element. It admits a left inverse that is a quasi-isomorphism: averaging over all pullbacks by group elements with respect to the Haar measure on \(G\).}
In that case, suppose that one has a sigma-model (not necessarily topological) whose configuration space consists of all smooth maps \(M\to G\). Then, a smooth map \(M\to G\) automatically induces a map \(\CE(\mathfrak g)\to \Omega(M)\), by taking the rational homotopy class. That is, one has a morphism of theories from the \(G\)-valued sigma model to the AKSZ-type theory valued in \(\mathfrak g\). Concretely, the geometric realization gives differential forms on \(G\) that realize \(\CE(\mathfrak g)\) (or equivalently, a subalgebra of the algebra of differential forms \(\Omega(G)\) isomorphic to the cdg-algebra \(\CE(\mathfrak g)\)). Then (the rational homotopy class of) a continuous map \(\phi\colon M\to G\) defines a field configuration of the AKSZ-type theory via pullback from \(G\).

When \(\CE(\mathfrak g)\) is minimal Sullivan, this morphism is injective up to rational homotopy, and also essentially surjective, in the sense that every cdg-homomorphism \(\CE(\mathfrak g)\to \Omega(M)\) is representable as a real linear combination of (homotopy classes of) continuous maps \(M\to G\). That is, it may be said to be a homotopic “skeleton” of the full sigma-model, classifying its topologically nontrivial configurations (skyrmions) that are not torsion. In effect, \(G\) can be thought of as a “universal solution” of the equations of the AKSZ-type theory, in the sense that every solution of the equations of motion of the AKSZ-type theory is realizable as a linear combination of those given by continuous maps \(X\to M\).


We also remark that if we wish to think of the \(X\)-valued sigma-model as the topological sector of a propagating quantum field theory, then \(X\) must have at least a smooth structure. However, if we are considering topological theories as independent entities in their own right, then there is no need to require a smooth structure on \(X\); a topological space (or an even more abstract homotopical notion) suffices.

\begin{example}
For example, when \(\mathfrak g\) is a Lie algebra, this inclusion is just the dual of the well known inclusion of \(\mathfrak g\) as the subalgebra of left-invariant vector fields: that is, the isomorphism, constructed in Lie algebra cohomology, between the Chevalley–Eilenberg chain complex and the invariant differential forms on \(G\).

In this case, the theory morphism discussed above reduces to the fact that, for the trivial principal bundle (say), every smooth map \(M\to G\) defines a \(\mathfrak g\)-valued form by pullback of invariant differential forms. However, since \(\CE(\mathfrak g)\) is not (minimal) Sullivan (that is, due to the presence of degree-1 elements), the “essential surjectivity” does not apply.
\end{example}

\begin{example}[6d continued]
The 6d AKSZ-type theory without scalars, 1-forms, and quadratic terms in the potential, with \(N\) 2-forms and \(N\) 3-forms, where the structure tensor \(\alpha_{ijk}\) is diagonal, is the Sullivan minimal model of the compact manifold \((\mathbb S^2)^{\times N}\). That is, its states classify the non-torsion skyrmions of the sigma model on this manifold.
\end{example}

Assuming the existence of a suitable geometric realization has far-reaching consequences: we are able to draw from and translate powerful topological theorems and mechanically translate them into statements about physics. For example,
\begin{itemize}
\item cdg-algebra morphisms that preserve the symplectic structure (“symplectomorphisms”) correspond to morphisms between AKSZ-type theories, at least classically: a classical solution to the domain theory can be transformed into a classical solution to the codomain theory.
\item the generators of the (torsion-free part of) homotopy groups correspond to generators of the Sullivan resolution.
\end{itemize}
In this way we are able to speak of the homotopy groups (i.e.~the space of fields of a given degree) or morphisms between AKSZ-type theories.

We give two examples of topological statements with physics translations.

\subsubsection{Elliptic–hyperbolic dichotomy}
Consider the condition that the geometric realization be (a) simply connected and (b) such that its total cohomology is finite-dimensional. These anodyne-seeming conditions turn out to produce powerful constraints on the number of generators in minimal Sullivan models. Let \(N\) be the degree of the highest-degree cohomology class.

Specifically, under the given condition, exactly one of the two conditions below must hold \cite{hess,fht}:
\begin{itemize}
\item Elliptic case: there are a finite number of generators; we will label the degrees of the odd generators as \(2a_1-1,2a_2-1,\dotsc, 2a_p-1\), and those of the even generators as \(2b_1,2b_2,\dotsc,2b_q\). Then the following must hold:
\begin{align}
\sum_jb_j - \sum_ia_i
&= \frac12(N + q - p)
&q &\le p \\
\sum_ia_j & \le \frac12N &
\sum_jb_i & \le \frac12(2N-1 + q)
\end{align}
\item Hyperbolic case: there are infinitely many generators; let \(d_i\) be the number of generators in degree \(i\). Then the following hold:
\begin{itemize}
\item the number of generators increases at least exponentially: \(\sum_{i\le n}d_i =\Omega(\exp(Cn))\) for some constant \(C>0\).\footnote{Here \(f(n) = \Omega(g(n))\) means \(\liminf_{n\to\infty} (f(n)/g(n)) > 0\).}
\item there are no long gaps, in thes sense that for every \(i\), there exists a \(j\) in the range \(i<j<i+N\) such that \(d_i >0\).
\end{itemize}
\end{itemize}
In other words, geometric realizability as a topological sigma model on a finite-dimensional topological space (not necessarily even by a compact a manifold!) implies complicated numerological constraints on the number and degrees of fields, at least if one excepts scalar and Yang–Mills fields.

\subsubsection{Lusternik–Schnirelman category}
Also, there is the rational Lusternik–Schnirelman category \(\operatorname{cat}_0(-)\) (see~e.g.~\cite{hess,clot}). In topology, the Lusternik–Schnirelman category of a topological space is the smallest possible size minus one\footnote{This normalization makes more sense for pointed spaces; older literature uses a different normalization without the \(-1\).} of an open covering of a space consisting solely of contractible open subsets; the \emph{rational} LS-category is the minimum LS-category of spaces rationally homotopy-equivalent to a given spaces. Now, there exists an equivalent definition of this invariant, as follows.
\begin{defn}
Given a minimal Sullivan cdg-algebra \(\bigwedge V\) and an integer \(n\ge2\), let \((\bigwedge V)^{>n} = (\bigwedge V) / \bigwedge^{>n}V\) be the cdg-algebra quotient with respect to the ideal \(\bigwedge^{>n}V\) consisting of words of length strictly greater than \(n\). Let the quotient map be
\[q_{>n}\colon \bigwedge V \to \frac{\bigwedge V}{\bigwedge^{>n}V}.\]
Now, there always exists a factorization of \(q_{>n}\) in the form
\[\bigwedge V \to \bigwedge (V\oplus W_n) \to \frac{\bigwedge V}{\bigwedge^{>n}V}\]
where \(\iota_n=\bigwedge V \hookrightarrow\bigwedge (V\oplus W_n)\) is an inclusion into a minimal Sullivan algebra, and \(\bigwedge (V\oplus W_n)\to (\bigwedge V)/(\bigwedge^{>n}V)\) is a surjective quasi-isomorphism. Now, we say that
\[
\operatorname{cat}_0\left(\bigwedge V\right) \le n
\]
iff \(\iota_n\) admits a left inverse
\[\rho_n \colon \bigwedge (V\oplus W_n) \to \bigwedge V,\]
that is, such a cdg-algebra homomorphism such that \(\rho_n\circ\iota_n \)  is the identity.
\end{defn}
Now, this definition can be interpreted line-by-line in physics terms:
\begin{itemize}
\item \(q_{>n}\) is the operation of applying the constraint where the product of more than \(n\) fields must vanish.
\item Such a constrained theory admits a Sullivan resolution, that is, an unconstrained “UV-completion”, such that the Abelian gauge moduli remain the same, and such that it includes the original unconstrained theory as a subtheory.
\item Now, we ask whether the additional set of fields \(W_n\) needed to implement this constraint is “trivial”, in the sense that there is a theory morphism where the additional fields \(W_n\) are simply replaced by suitable products of the original fields \(V\). 
\end{itemize}
Now, the problem with this interpretation is that in general there is no invariant quadratic form on \(V\oplus W\) (i.e.~does not admit a Lagrangian formulation), and may in fact involve “fields” of degree higher than the manifold.


The latter is monotonic in the sense that if a morphism \(X\to Y\) that is injective in the homotopy group exists, then
\[\operatorname{cat}_0(X) \le \operatorname{cat}_0(Y).\]
That is, it is some measure of the “complexity” of a theory. It is furthermore additive \cite[Theorem~2.19]{hess}:
\[\operatorname{cat}_0(X\times Y) = \operatorname{cat}_0(X) + \operatorname{cat}_0(Y).\]
In terms of physics, the product-space \(X\times Y\) sigma model amounts to having two sets of fields, not interacting with each other. That is, this measure is well behaved under the (trivial) “product” of theories. (The disjoint union or wedge sum of spaces instead corresopnds to superselection sectors, i.e.~direct sum of the respective Hilbert spaces instead of tensor product.)

In particular, the rational Lusternik–Schnirelman category can be easily computed, and its computation can be given physics interpretations.

\section{Conclusion}
This paper contains an expos\'e of the intimate relation between a certain class of algebras arising in rational homotopy theory and categorical generalizations of Lie algebras on one hand and a certain class of topological quantum field theories on the other hand.

We again note that we have limited ourselves in this paper to the consideration of the local and classical aspects of the theory: local, in the sense that we have neglected the possible nontrivial topology and geometry of the gerbes of which the fields are flat connections; classical, in the sense familiar to physicists. We hope to address these issues in upcoming papers.

One natural avenue of further exploration is to consider more general types of physics actions, which should dualize to more restricted types of spaces equipped with additional structure. The next simplest types of physical theories after topological ones are those that depend on a volume form, so that e.g.~Hodge stars can occur in the action. Dually, this class of theories should correspond to topologies equipped with additional structure that defines this pairing.

Another avenue, more mathematical in nature, is to discuss more systematically the structure of the categories of these algebraic structures. Several natural operations, including the Sullivan resolution of non-simply-connected quasifree cdg-algebras, do not naturally respect the symplectic structure. On the other hand, operations more directly motivated from physics, such as dimensional reduction, do naturally preserve it.

Another consideration is the relation of AKSZ-type theories with the identification of \(\Omega(M)\) as the Hilbert space of a differential-form quantum mechanics models. The paper \cite{ks} discusses types of models, and gives a physical interpretation of Sullivanization in that context.

\bibliographystyle{alphaurl}
\bibliography{biblio}

\end{document}